\RequirePackage{lineno}
\documentclass[fleqn,10pt]{wlscirep}
\usepackage[utf8]{inputenc}
\pdfoutput=1
\usepackage[T1]{fontenc}
\usepackage{siunitx} 
\DeclareMathOperator*{\argmax}{arg\,max}
\DeclareMathOperator*{\argmin}{arg\,min}
\title{Ultrasonic Actuation of a Fine-Needle Improves Biopsy Yield}

\author[1]{Emanuele Perra}
\author[2]{Eetu Lampsijärvi}
\author[3,4]{Gonçalo Barreto}
\author[1]{Muhammad Arif}
\author[2]{Tuomas Puranen}
\author[2]{Edward Hæggström}
\author[5,6]{Kenneth P.H. Pritzker}
\author[1,*]{Heikki J. Nieminen}

\affil[1]{Medical Ultrasonics Laboratory (MEDUSA), Department of Neuroscience and Biomedical Engineering, Aalto University, Espoo, 02150 , Finland}
\affil[2]{Electronics Research Laboratory, Department of Physics, University of Helsinki, Helsinki, 00560, Finland}
\affil[3]{Translational Immunology Research Program, University of Helsinki, Helsinki, 00100, Finland}
\affil[4]{Orton, Helsinki, 00280, Finland}
\affil[5]{Department of Laboratory Medicine and Pathobiology, University of Toronto, Toronto, M5S 1A8, Canada}
\affil[6]{Department of Pathology and Laboratory Medicine, Mount Sinai Hospital, Toronto, M5G 1X5, Canada}
\affil[*]{heikki.j.nieminen@aalto.fi}

\makeatletter
\renewcommand{\@maketitle}{%
{%
\thispagestyle{empty}%
\vskip-36pt%
{\raggedright\sffamily\bfseries\fontsize{20}{25}\selectfont \@title\par}%
\vskip10pt
{\raggedright\sffamily\fontsize{12}{16}\selectfont \@author\par}
\vskip25pt%
}%
}%
\makeatother

\begin{document}

\flushbottom
\maketitle
% Click the title above to edit the author information and abstract
%
\thispagestyle{empty}
\section*{ABSTRACT}
Despite the ubiquitous use over the past 150 years, the functions of the current medical needle are facilitated only by mechanical shear and cutting by the needle tip, \textit{i.e.} the lancet. In this study, we demonstrate how nonlinear ultrasonics (NLU) extends the functionality of the medical needle far beyond its present capability. The NLU actions were found to be localized to the proximity of the needle tip, the SonoLancet, but the effects extend several millimeters from the physical needle boundary. The observed nonlinear phenomena, transient cavitation, fluid streams, translation of micro- and nanoparticles and atomization, were quantitatively characterized. In the fine-needle biopsy application, the SonoLancet contributed to obtaining tissue cores with increase in tissue yield by 3-6x in different tissue types compared to conventional needle biopsy technique using the same 21G needle. In conclusion, the SonoLancet could be of interest to several other medical applications, including drug or gene delivery, cell modulation, and minimally invasive surgical procedures.

\section*{Introduction}

The antecedent of the medical lancet was described by Hippocrates for the purpose of puncturing and draining pus\cite{Kirkup1986}. Currently the lancet, a small lance\cite{Bynum2015}, is widely employed in medical cutting and piercing devices, \textit{e.g}. hypodermic needles, surgical tools and tissue sampling needle biopsy devices. Structurally a lancet typically has at least two converging edges and a sharp tip\cite{Kucklick2005}. When pressed against or moved along a tissue interface, the lancet’s purpose is to utilize manually mechanical forces to separate one structure from another in order to achieve or enable a medical purpose\cite{Wang2013}. The medical needle, employing lancet-shape or other needle tip geometries\cite{Han2012}, is a common tool in healthcare, exemplified by the estimated 16 billion annual injections worldwide\cite{Hauri2004}. However, the functions of the needle tip are limited to mechanical shear and cutting by the needle tip. In fact, while not extensively studied, previous literature suggests that some needle functions still present limitations with regard to pain (about 10\% of the population suffers from needle phobia\cite{Hamilton1995} and 3-10 \% of the population has been estimated to avoid medical care because of the fear of needles \cite{Jenkins2014, AmericanPsychiatricAssociation2013, Ayala2009}), precision\cite{Rossa2016,Wan2005}, spatial localization\cite{Ji2007} and for needle biopsies, adequacy with regard to quality and quantity of tissue yield\cite{Pritzker2019}.

An example of a field limited by current functions of a medical needle is needle biopsy for cancer diagnosis. To provide a cancer therapy, histological and molecular tissue samples are required. However, up to 34\% of histological and up to 50\% molecular assessments fail, because of obtaining diagnostically insufficient sample characterized by limited quantity of obtained pathological cells or nucleic acids\cite{Pritzker2019,Gutierrez2017}. Insufficiently adequate biopsies induce a burden on the patients due to extended wait times before obtaining the diagnosis\cite{E2005}. The time pressure to start the therapy can lead to the treatment initiation with inadequate information. Therefore, ways to improve tissue yield, while not compromising safety, are urgently needed in cancer diagnostics.

Actuation of medical needles by ultrasound has been observed to induce tissue movement that can help with localizing the needle under Doppler ultrasound\cite{Gui2014,Kuang2016} or to reduce penetration resistance of the needle\cite{Liao2014a,Liao2013}, which could potentially help in reducing the pain of the needle insertion through the skin \cite{Kundan2019, Kaushik2001}. In conjunction with microneedles, low frequency mechanical oscillations in the kHz range have been used for oocyte micro-dissection\cite{Gao2018}, to enhance the microneedle penetration into mouse embryos\cite{Bahadur2017}, fish eggs\cite{Wu2009} and oocytes\cite{Johnson2018}, and subsequently to improve oocyte viability for in vitro fertilization\cite{Yanagida1999}. However, until now, the potential of NLU extending beyond the needle tip has been overlooked as a way to add value to the function of medical needles. In fact, the NLU can generate acoustic force fields providing a precise manipulation of entities selectively closer or at a farther distance from the sound source. 

Fine-needle aspiration biopsy (FNAB) is a common biopsy method in which a hypodermic needle and a syringe are employed to aspirate tissue constructs from a target tissue, \textit{e.g.} suspected tumor. The needle is translated by the operator, while the needle tip is inside the target tissue. Suction by the syringe causes the tissue to bulge towards the needle lumen and adhere to the inner walls of the needle, while the rapid movement of the needle tip will cause the protruded portion of tissue to be sliced off by the cutting edges of the needle and to be aspirated into the lumen\cite{Kreula1990}. Here the source of the external force is the operator hand, which translates to compressive, shear and tensile forces at the needle tip and explain the detachment of cells and tissue constructs from the target tissue. Contrary to FNAB, we hypothesize that ultrasound could serve as an external source of forces near the needle tip further accelerating the detachment of cells and tissue constructs. 

Coupling flexural waves to a needle induces sideways motion of the needle tip. This is anticipated to exert direct shear, compression and tensile stresses to the needle tip environment as well as induce sound emission, beyond the forces induced by the operator’s hand movement. The sound emission could generate NLU phenomena such as acoustic radiation force\cite{Dayton1999}, acoustic streaming\cite{Nyborg1958} and cavitation\cite{Neppiras1951}. In the context of biopsy, the direct and indirect forces are expected to contribute to extracting cells and tissue constructs, which could enhance the yield in biopsy.

The aim of this study was to investigate how adding ultrasound to a conventional medical needle could change its conventional functions. We first demonstrated how ultrasound emission could be localized at a tip of a conventional medical needle, as an energy source for NLU around the needle tip. From micrometer to millimeter scale, we quantitatively characterized the NLU effects potentially contributing to tissue actuation, such as cavitation, acoustic radiation force and atomization spatially reaching beyond the physical lancet; we defined the confined volume around the needle tip exhibiting pronounced nonlinear effects as the \textit{SonoLancet}. Finally, we demonstrated the capability of SonoLancet to increase the sample yield in the FNAB application.

\section*{Results}

\subsection*{Localization of ultrasound near the needle tip}
To demonstrate that the effects potentially contributing to tissue actuation can be generated near an ultrasonically actuated medical needle (Figure \ref{Figure1}A(1-3)), a 21 gauge hypodermic needle was selected, because it represents a common medical needle. Moreover, needles of this diameter or less (fine needles) induce less tissue trauma than larger needles. The selected geometry permits access to the needle hub with a variety of adjuvant devices commonly used in medical practice, \textit{e.g}. tubing, syringe, vacutainer; for this study we selected a \SI{10}{\milli\liter} syringe. 

We employed an axially translating Langevin transducer as a sound source, which was coupled to an S-shaped waveguide connected to a medical needle (Figure \ref{Figure1}A(1)). Using the waveguide, the longitudinal mode waves of the ultrasound transducer are converted to flexural mode waves in the needle. Reflection of the wave from the needle tip allowed generation of a flexural standing wave between the tips of the needle and the waveguide. Moreover, the converging structure of the medical needle bevel gives geometric amplification of the wave towards the needle extremity. As a consequence of the wave amplification, the needle tip is made to act as a dipole-like sound source oscillating at a large displacement amplitude (Figure \ref{Figure1}D), exhibiting considerably greater time-averaged acoustic intensity at the needle tip (Figure \ref{Figure1}B) than elsewhere near the needle lumen, which carries the energy. These characteristics have specific relevance to non-linear ultrasonics (NLU) associated with time-averaged intensity (\textit{e.g}. acoustic radiation pressure) or characterized by threshold behavior (\textit{e.g}. cavitation), since the NLU phenomena can be limited to the proximity of the needle tip.

As the ultrasound field near the needle is largely confined within a distance of few mm near the needle bevel (Figure \ref{Figure1}B(1)), the localization of the SonoLancet is spatially precise and microtrauma in tissues farther away from the needle tip is expected to be minimized. In fact, the threshold phenomenon of cavitation and the fast geometric attenuation of the acoustic field near the needle exclude explanations, where cavitation could unexpectedly occur very far from the needle. Cells and tissue components influenced by the SonoLancet can be withdrawn through the needle lumen directly from the site of ultrasound action or alternatively, cells and agents can be delivered to a spatially well-defined site in conjunction with ultrasonic actions.

\subsection*{Cavitation}
\begin{figure}[ht]
\centering
\includegraphics[width=\linewidth]{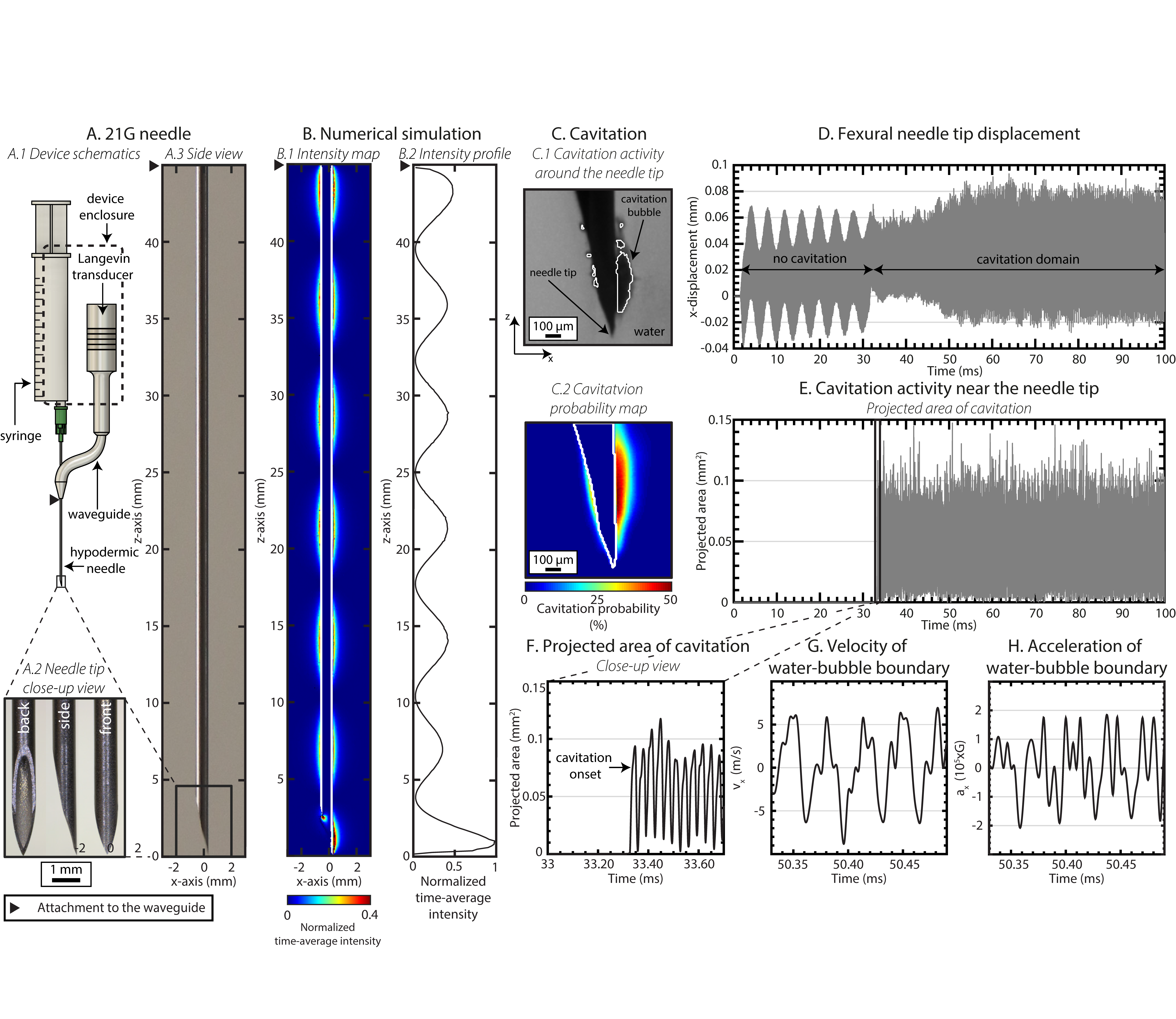}
%\internallinenumbers
\caption{(A(1)) Schematic represents the investigational device that comprises a conventional Langevin ultrasound transducer coupled \textit{via} an aluminum waveguide to a (A(2-3)) 21G hypodermic needle connected to a \SI{10}{mL} syringe. The geometry permits connecting virtually any pressure source to the needle. (B(1)) Numerical simulation result representing the x-component of the time-average acoustic intensity evaluated on the xy-plane coincident with the needle center axis and on a line adjacent to the outer needle surface, starting from the needle tip and ending at the waveguide-needle attachment. These results demonstrate localization of sound energy at the very tip of the needle, more than double the intensity calculated elsewhere. (C(1)) The cavitation activity is highly concentrated at the needle tip, as supported by (C(2)) the projected spatial probability of cavitation in deionized water ([$O_2$] \SI{5.8}{\mg\per\L}). (D,E,F) Temporally we observed a cavitation onset after which the inertial cavitation events continued in an uninterrupted manner. (G) The applied ultrasound induced peak velocities up to \SI{5}{\m\per\s} and (H) acceleration of the bubble and water boundary that was equivalent to \SI{20000}{G}. The results demonstrate that a conventional medical needle can be converted into a highly controlled, ultrasonically functionalized instrument with significant NLU phenomena concentrated at the very tip of the needle.}
\label{Figure1}
\end{figure}

Large displacements in the direction of x-axis (Figure \ref{Figure1}D) detected near the lancet point produce pronounced acoustic intensity radiating outwards from the center axis of the needle. The high magnitude of the peak-negative-pressure, associated with the acoustic intensity, elevates the probability of ultrasound-microbubble interactions, \textit{i.e.} cavitation (Figure \ref{Figure1}C(1,2)). Cavitation activity exhibited within a region extending less than \SI{2}{\mm} from the tip along the positive z-direction. Optical high-speed (HS) imaging (\SI{300000}{fps}) revealed that the temporal probability of observing cavitation across a time window of \SI{100}{\ms} can be up to 50\% in this region (Figure \ref{Figure1}C(2)), while in the proximity of the rest of the needle there were no signs of cavitation. This spatio-temporal behaviour arises from the inertial cavitation being a threshold phenomenon\cite{Holland1990}. Cavitation, \textit{i.e.} growth, oscillation and collapse of gas bubbles, is a consequence of the fluctuating pressure amplitude being momentarily the lowest in this area due to the pronounced amplification of needle tip motion compared to other locations along the needle lumen (Figure \ref{Figure1}B(1,2)). 

While the needle tip displacement was observed to be < \SI{100}{\micro\meter} (Figure \ref{Figure1}D), the bubble-water boundary displacement was more pronounced, extending beyond \SI{300}{\micro\meter} along the positive x-direction. This is due to the different compliances of water and an air bubble. The activity of the primary bubble within the SonoLancet produced fluid movements of several \SI{}{\meter\per\second} (Figure \ref{Figure1}G), accelerations up to five orders of magnitude of gravity (Figure \ref{Figure1}H), thus allowing production of high shear stresses localized near the needle tip\cite{Collis2010}. Collapsing bubbles produce secondary sound formation, a potential mechanism of shock wave formation\cite{Brujan2005} and strong and transient shear forces have great potential to actuate matter, such as micro- and nanoparticles or tissue such as cells, groups of cells or localized regions of an organ. 

\subsection*{Acoustic radiation pressure}

\begin{figure}[ht]
	\centering
	\includegraphics[width=\linewidth]{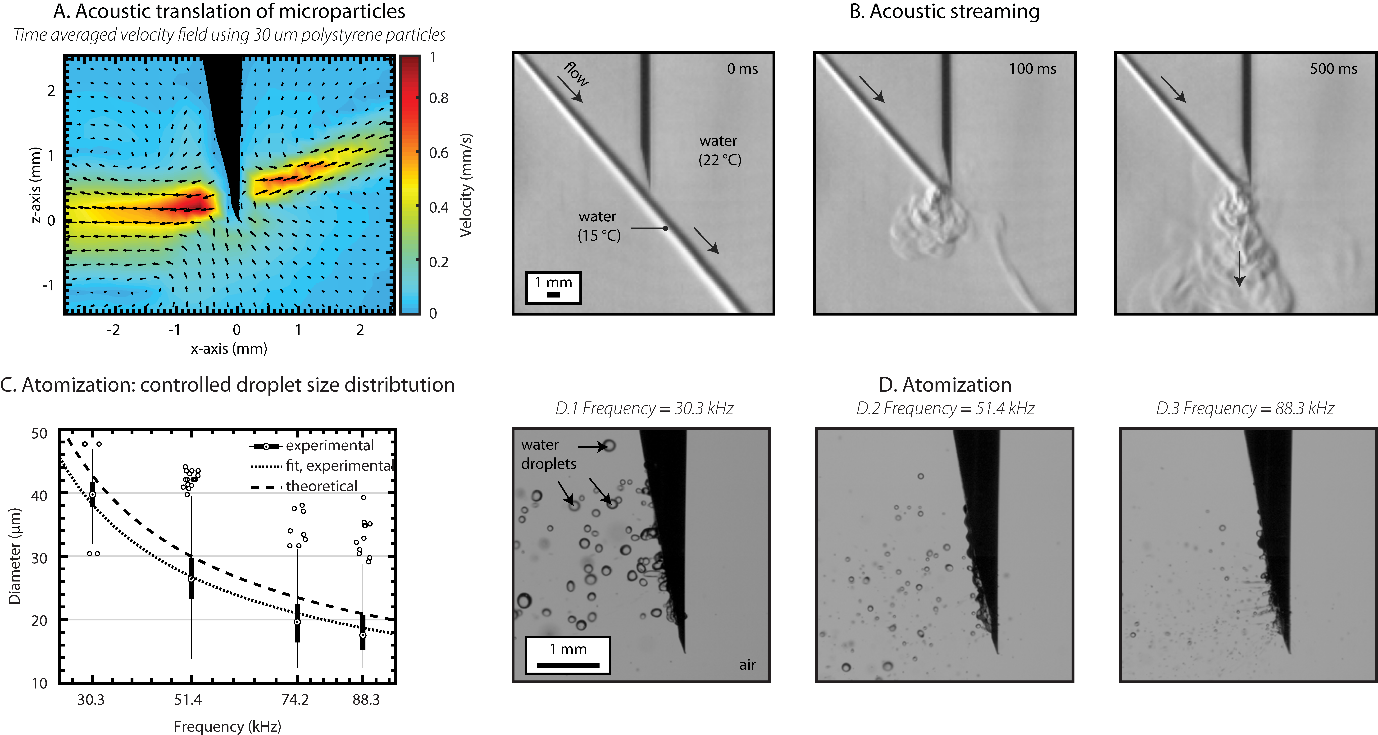}
	%\internallinenumbers
	\caption{(A) \SI{30}{\um} diameter polystyrene microparticles were used to visualize the water flow pattern around the needle during sonication. The graph represents the time-averaged velocity field of the particles, calculated across a time window of \SI{100}{ms}. (B) Schlieren images show the disturbance of water laminar flow induced by the needle action as a demonstration of acoustic streaming taking place around the needle tip during sonication. (C) The droplet size distribution was controlled with ultrasonic driving frequencies and followed the theoretical relationship. This is exemplified in HS imaging (D), which revealed micro-droplet ejections of water at different driving frequencies from a main water drop hanging from the needle when the needle was embedded in air. (A, B) The results demonstrate the capability of SonoLancet to induce mass transfer as well as (C, D) controlling the size of atomized droplets. On each box, the central mark represents the median, and the lower and upper edges of the box indicate the 25th and 75th percentiles, respectively. The whiskers extend to the most extreme values of the data set excluding the outliers, and the outliers are plotted individually with non-filled circles.}
	\label{Figure2}
\end{figure}

Acoustic radiation force can be employed to push or pull medium, objects and interfaces\cite{Sapozhnikov2013}. It is a force exerted on a target arising from acoustic radiation pressure, which is due to a change in momentum of the acoustic wave, when the wave interacts with such target. We here studied the capability of SonoLancet to generate acoustic radiation pressure in a suspension of microparticles. Two main jets (Figure \ref{Figure2}A) of microparticles (models for micro-vehicles), were emanating from opposite sides of the needle bevel with a maximum velocity of approximately \SI{1}{\mm\per\second} at \SI{0.6}{\mm} from the needle tip, decreasing \SI{0.5}{\mm\per\second} at a distance of \SI{2}{\mm} mm away from the maximum. Considering the wavelength, the small object size and similar acoustic impedance of particles compared to that of water, the main mechanism of translation arises from acoustic streaming, \textit{i.e.} acoustic radiation force exerted on the liquid, rather than acoustic radiation force directly pushing the particles. This is supported by a finding that SonoLancet in water (\SI{22}{\celsius}) without microparticles induced acoustic streaming; such observation was confirmed under Schlieren imaging (Figure \ref{Figure2}B), which revealed a disturbance of water laminar flow (\SI{15}{\celsius}) close to the sonicating needle. The results demonstrate that SonoLancet is capable of inducing acoustic radiation pressure leading to mass transfer near the needle tip, which could be relevant to delivery of directional transportation of liquids or particles in localized therapeutic applications such as gene and drug delivery.

\subsection*{Atomization}

Capillary waves propagating at the water-air interface can induce micro-droplet formation at large displacement amplitudes\cite{Vukasinovic2007}. To investigate the potential of SonoLancet to produce micro-droplets, we first introduced a water droplet near the needle tip, while the needle was embedded in air. Capillary waves were observed at low driving power levels (electrical output power << \SI{1}{W}) of the transducer. At increased powers (electrical output power $\sim$ \SI{1}{W}) crests of the waves started to extrude, eventually forming droplets ejected from the water-air interface\cite{Vukasinovic2007}. By adjusting the driving frequency, we modified the size distribution of the deionized water droplets ejected from the surface (Figure \ref{Figure2}C-D) from a median size of \SI{39.5}{\um} down to \SI{17.5}{\um}. The trend in median droplet size is in line with the theoretical prediction:
%\begin{linenomath*}
\begin{equation}
\begin{aligned}
d_p = 0.34\Big(\frac{8\pi\sigma}{\rho f^2}\Big)^{1/3}
\end{aligned}
\label{lang}
\end{equation}
%\end{linenomath*}
 where $d_p$ is the droplet size, $\sigma$ is the surface tension, $\rho$ is the density and $f$ is the excitation frequency\cite{Rajan2001}. These results demonstrate the capability of SonoLancet to produce water micro-droplets with a controlled size distribution, relevant to \textit{e.g}. delivery of therapeutic agents such as drugs or antiseptics in pulmonary conditions.

\subsection*{Influence of SonoLancet on a tissue phantom and tissue}

\begin{figure}[ht]
	\centering
	\includegraphics[width=\linewidth]{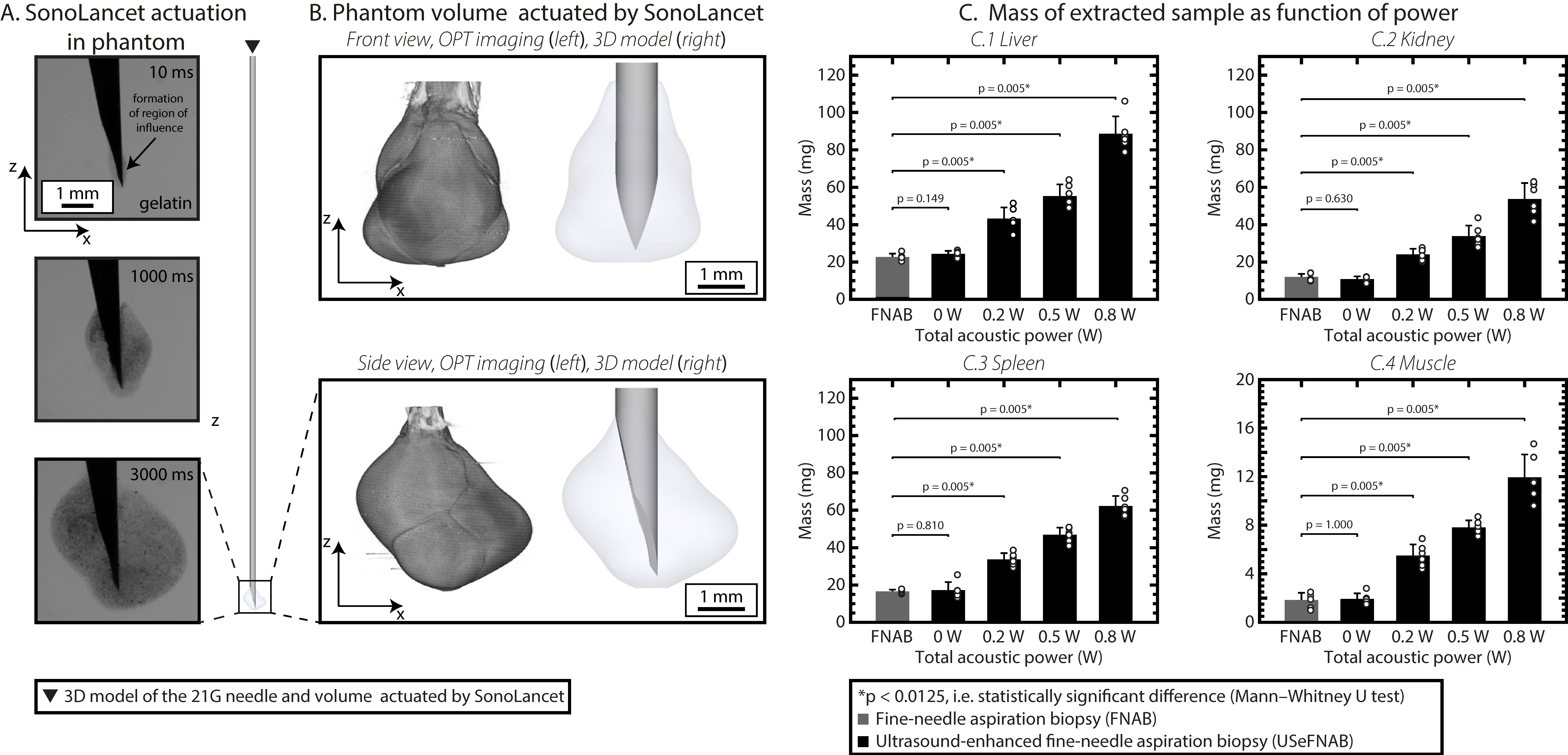}
	%\internallinenumbers
	\caption{(A) HS imaging of experiments in tissue phantom (10\% w/v ballistic gelatin/deionized water) reveals projections of the volume of interaction between the gelatin and the Sonance at different time points during continuous sonication. (B) A 3D reconstruction of the actuated region was produced by examining the sample under optical projection tomography. (C) The mass of extracted sample in ultrasound enhanced biopsies performed in bovine (C(1)) liver, (C(2)) kidney, (C(3)) spleen and (C(4)) muscle at different power levels compared to the standard FNAB. The needle employed in FNAB and ultrasound-enhanced FNAB (USeFNAB) was the same 21G needle. In the bar charts, the bar height represents the mean of the data set and the error bar indicates the standard deviation. Individual data points are shown as white-filled circles.}
	\label{Figure3}
\end{figure}

As demonstrated above, the SonoLancet can induce nonlinear effects in well-defined media such as water and air. In the following, we demonstrate SonoLancet’s influence on a tissue phantom and on excised tissue. Ballistic gelatin was selected as a model for visualizing the needle activity inside a soft-tissue-like material. We subjected a hydrogel tissue phantom (10\% w/v porcine gelatin, \SI{22}{\celsius}) to the SonoLancet in conjunction with fluorescent nanoparticles (diameter \SI{63}{nm}) introduced to the proximity of the needle tip through the lumen. When a continuous wave of \SI{31.7}{\kHz} and total acoustic power (TAP) $\sim$ \SI{0.8}{W} were applied to the needle, the generation of cavitation bubbles and acoustic streaming was optically confirmed. The boundary of the influenced volume extended outward from the needle tip over time (Figure \ref{Figure3}A). After sonication was applied together with fluorescent nanoparticles, a 3D reconstruction of the nanoparticle distributions characterized by optical projection tomography revealed a volume containing nanoparticles is localized near the needle tip (Figure \ref{Figure3}B). Dissection of the phantom was used to verify that the gel was liquefied. At the highest TAP levels, $\sim$ \SI{0.8}{W}, we recorded a temperature rise of < \SI{2}{\celsius} or < \SI{4}{\celsius} during a 5 or \SI{20}{\second} sonication, respectively, suggesting that cavitation was the main mechanism of liquefaction\cite{Bader2018}, rather than ultrasound-induced temperature rise. Even though the observed acoustic phenomena in gelatin may occur in a different fashion in an ex vivo tissue, this experiment gave us an insight on how the acoustic energy is manifesting an influencing the surroundings of the needle tip, when the needle has been confined by material with similar acoustic impedance to that of soft tissue.

\begin{figure}[ht]
	\centering
	\includegraphics[width=\linewidth]{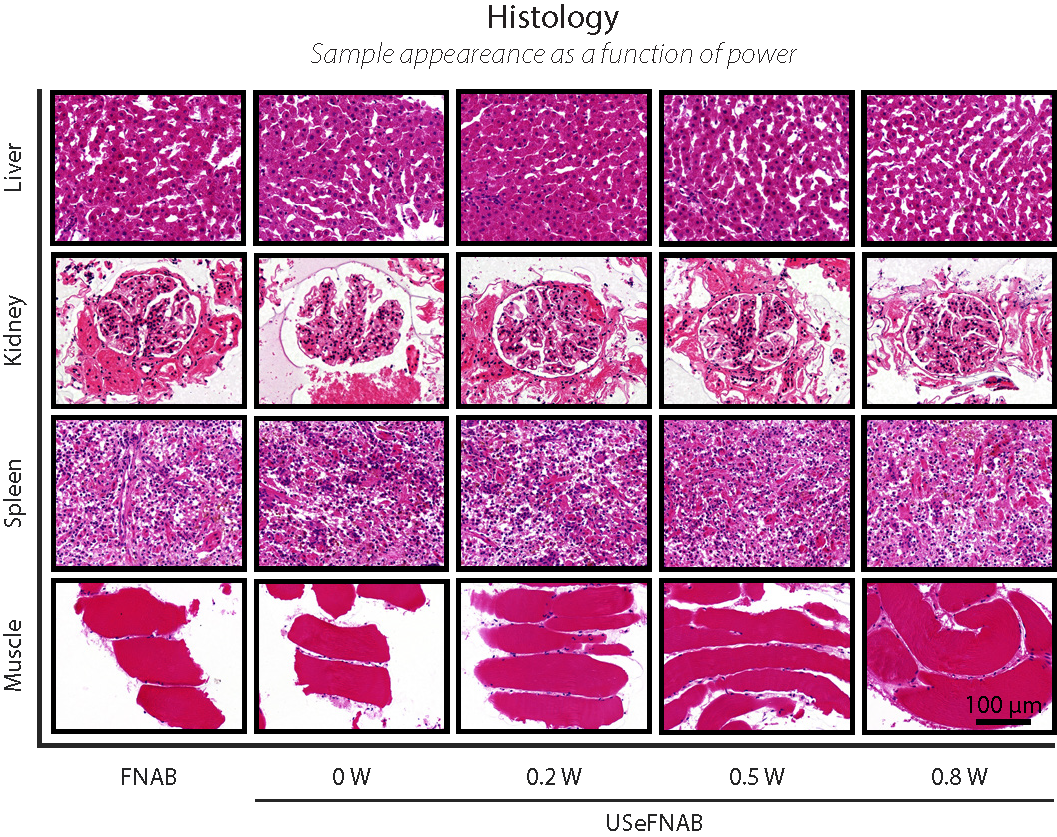}
	%\internallinenumbers
	\caption{Histological pictures (20x) of cell block sections prepared from biopsy aspirates taken from different bovine organs (top to bottom: liver, kidney, spleen and muscle) at different TAP levels (left to right: FNAB, USeFNAB 0 W, 0.2 W, 0.5 W and 0.8 W). The results demonstrated that different tissue types included histopathologically relevant cells or tissue constructs, suggesting that SonoLancet could allow histopathological evaluation of samples in selected FNAB applications, if developed to its full potential.}
	\label{Figure4}
\end{figure}

To demonstrate the influence of SonoLancet on tissue, we performed FNABs with and without ultrasound at the needle tip (ultrasound-enhanced fine-needle aspiration biopsy, USeFNAB) in different bovine tissues: liver, kidney, spleen and striated muscle. USeFNAB contributed to increased sample yield, whose weight increased with increasing TAP (\textit{p} < 0.0125), being on average up to 4x the yield obtained with FNAB in liver, 5x in kidney, 3x in spleen and 6x in muscle (Figure \ref{Figure3}C), as compared to FNAB. When ultrasound-enhanced biopsies were performed at the highest TAP (\SI{0.8}{W}), the average sample weight obtained in liver was \SI[multi-part-units = single]{88.6 (93)}{mg} (average $\pm$ standard deviation; \textit{n} = 6) (Figure \ref{Figure3}C(1)), \SI[multi-part-units = single]{53.7 (86)}{mg} (\textit{n} = 6) in kidney (Figure \ref{Figure3}C(2)), \SI[multi-part-units = single]{62.2 (54)}{mg} (\textit{n} = 6) in spleen (Figure \ref{Figure3}C(3)) and \SI[multi-part-units = single]{11.9 (19)}{mg} (\textit{n} = 6) in muscle (Figure \ref{Figure3}C(4)); the sample masses obtained with FNAB were respectively \SI[multi-part-units = single]{22.7 (18)}{mg} (\textit{n} = 6), \SI[multi-part-units = single]{11.7 (20)}{mg} (\textit{n} = 6), \SI[multi-part-units = single]{16.6 (10)}{mg} (\textit{n} = 6) and \SI[multi-part-units = single]{1.8 (6)}{mg} (\textit{n} = 6). Based on a histopathological evaluation, performed by a pathologist (KP) with > 40 years of histopathology and cytopathology experience, all tissue types included intact cells and tissue (Figure \ref{Figure4}) containing histologically relevant structures that could be observed in all samples. 

\begin{figure}[ht]
	\centering
	\includegraphics[width=\linewidth]{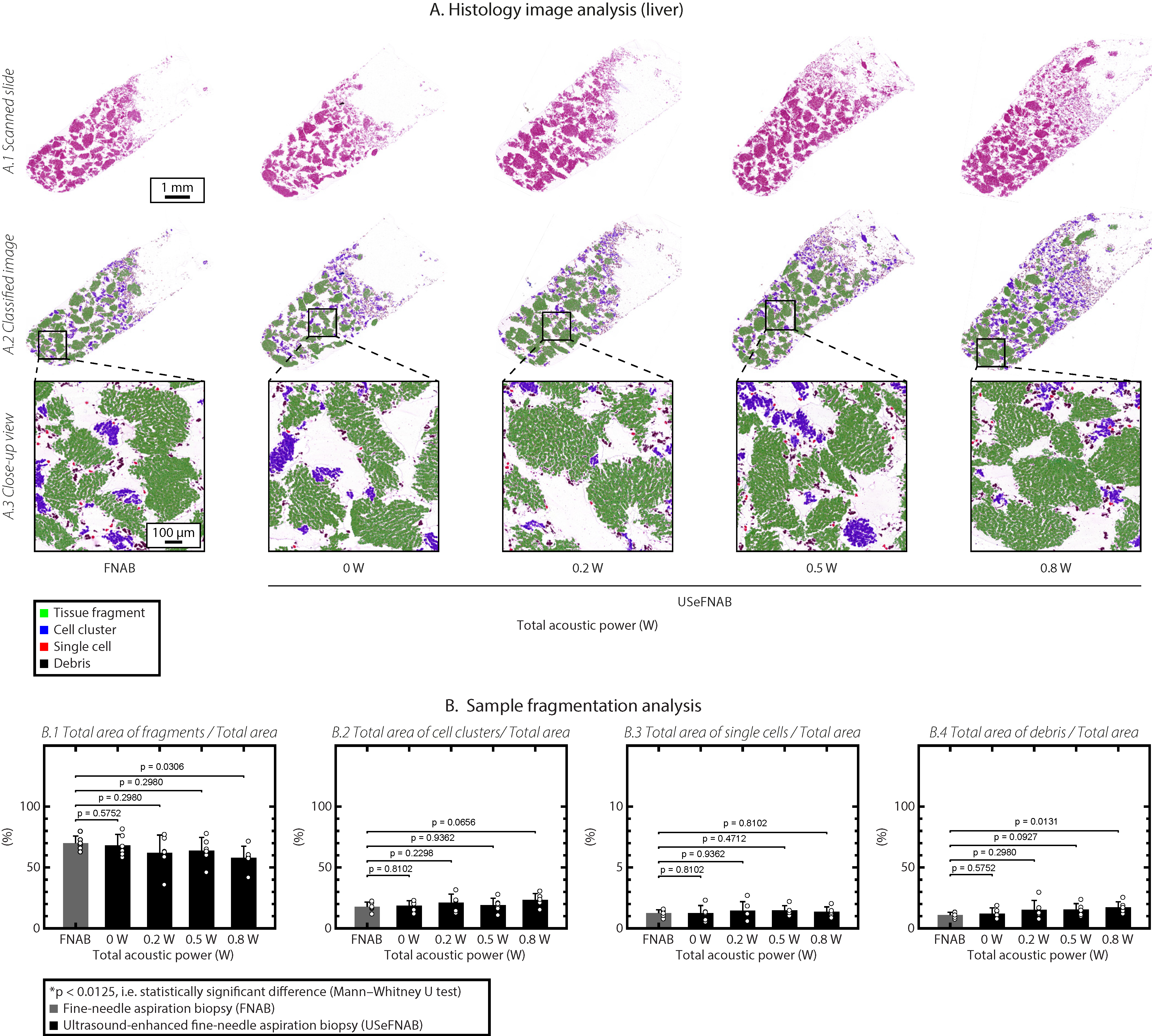}
	%\internallinenumbers
	\caption{(A) Computer assisted evaluation of tissue fragmentation in liver biopsy samples. (A(1)) represents exemplary histological images used for the analysis. (A(2,3)) show the classified images where tissue fragments, cell clusters, single cells and debris are detected and highlighted in green, blue, red and black, respectively. (B(1)) The relative total area of tissue fragments, (B(2)) cell clusters, (B(3)) single cells, (B(4)) debris pieces were calculated and used for statistical comparison between the FNAB and USeFNAB groups. In the bar charts, the bar height represents the mean of the data set and the error bar indicates the standard deviation. Individual data points are shown as white-filled circles.}
	\label{Figure5}
\end{figure}

A computer-assisted method was adopted to quantify the fragmentation in the liver samples (Figure \ref{Figure5}A). The area of tissue fragments (large regions of adjacent hepatocytes), cell clusters (small aggregation of hepatocytes), single cells and debris pieces (organic waste) detected in the control group slides (FNAB) were measured and compared to the experimental groups (USeFNAB, TAP levels: \SI{0}{W}, \SI{0.2}{W}, \SI{0.5}{W} and \SI{0.8}{W}). On average the total areas of tissue fragments, cell clusters, single cells and debris pieces were slightly influenced by ultrasound; however, they were not statistically significantly different from the respective areas of FNAB (\textit{p} > 0.0125). At the highest TAP the relative total area of tissue fragments was lower (\SI[multi-part-units = single]{57.9 (95)}{\%}, average $\pm$ standard deviation, \textit{n} = 6, p = 0.0306) as compared to FNAB (\SI[multi-part-units = single]{69.9 (58)}{\%}, \textit{n} = 5) (Figure \ref{Figure5}B(1)). The relative total area of cell clusters and individual cells were not different from those of FNAB when the highest TAP was employed (p = 0.0656 and p = 0.8102, respectively) (Figure \ref{Figure5}B(2,3)). The relative total debris area (\SI[multi-part-units = single]{17.2 (46)}{\%}, \textit{n} = 6, p = 0.0131) was higher as compared to FNAB (\SI[multi-part-units = single]{11.0 (23)}{\%}, \textit{n} = 5) (Figure \ref{Figure5}B(4)). 

The results show a systematic increase in the sample mass with increasing TAP, while inducing a minimal effect on the morphological structure of the sample as compared to the FNAB approach, as a manifestation of the actuation of tissue induced by SonoLancet.

\section*{Discussion}

We successfully coupled flexural standing waves to a conventional medical needle. The converging structure of the needle bevel seemed acoustically quite optimal at the employed frequency, because it geometrically amplified the wave allowing to generate a highly localized and large displacements of the needle tip. The acoustic intensity of the longitudinal waves emitted from the needle extended beyond the linear regime to nonlinear regime, suggested by large displacements < \SI{100}{\um} and the observed nonlinear effects. The maximum time-averaged acoustic intensity next to the needle bevel was more than twice compared to that along the lumen. Since the acoustic radiation pressure associated phenomena, such as acoustic radiation force or acoustic streaming, are proportional to intensity, the mass transfer mechanisms are pronounced at the needle tip. The long needle shaft has multitude of resonant eigen-frequencies, which allowed changing the operation frequency. This permitted controlling the size of the atomized droplets contrary to common ultrasound atomizers, which typically operate on a single frequency\cite{Barreras2002,James2003,Ueha1985}.

Cavitation being a threshold phenomenon allowed generation of cavitation at will. Importantly, this was spatially restricted to the proximity of the needle tip without observations of cavitation along the needle lumen, suggesting that these mechanisms can be made to localize to the proximity of the needle tip. The high accelerations > \SI{20000}{}x the gravitational acceleration of the bubble-water boundary suggests that extreme shear forces can be induced by this boundary to the surroundings, when cavitation is intended. In addition, the direct effect of the rapidly moving needle tip with sharp edges provide points to exert high stress fields on matter, such as tissue. 

Minimally invasive high-frequency mechanical actuation at a known position inside the body has specific relevance to various medical applications. We exemplified the influence of the SonoLancet on tissue in the FNAB application. Depending on the tissue type, the tissue yield was increased 3-6x on average and with increasing TAP. A potential explanation of the enhanced mass of the biopsy sample lies in the combined action of the localized movement of the needle tip and the suction force applied with the syringe. The suction force is needed to pull the tissue against the needle tip and to prevent it to slip during the cutting process. However, the suction alone does not tear cells or tissue fragments from the target. In fact, Kreula \textit{et al.} demonstrated that the sample yield is linearly dependent on the suction force only when combined with the needle movement (\textit{e.g.} fanning)\cite{Kreula1990}. Based on this finding, we believe that the ultrasonic actuation of the needle introduces an additional degree of freedom in the needle movement (flexural displacement of the needle tip) that introduces additional forces facilitating the tissue extraction in an analogous way to what the fanning technique would contribute to, but at a smaller length-scale and higher frequency. In addition, the studied NLU phenomena are expected to generate shear forces that contribute to the extraction of cells and tissue constructs from the target, which are then drawn into the needle by the low pressure applied with the syringe. We foresee this to have importance especially in cancer diagnostics, which increasingly relies on molecular assessments for therapy decisions. The failure rate of molecular assessment is up to 50\%\cite{Pritzker2019} arising mainly from inadequacy of the nucleic acids required for determining the cancer type. This could be resolved, by obtaining a greater sample volume using SonoLancet permitting access to greater nucleic acid quantity. The SonoLancet could contribute also to histopathological assessments, which suffer from failure rates up to 34\% due to inadequacy of tissue yield\cite{Carson1995,Nassar2011}.

Future studies should address the clinically relevant adequacy of the sample, its micro-level integrity and the influence of ultrasound on the needle tract.

SonoLancet could be also beneficial for other medical applications beyond cancer diagnostics. Controlled cavitation, used in a variety of existing and emerging ultrasonic medical technologies\cite{Brennen2015}, could be exploited in personalized therapeutic and diagnostic applications, by applying cavitation locally at the needle tip precisely at the target site. Such applications include tumor ablation\cite{Zhou2011}, permeation of tissue matrix for enhanced stem cell delivery or migration, bubble-enhanced poration of cells, \textit{i.e.} sonoporation\cite{Mehier-Humbert2005}, transfection of genetic material\cite{Chettab2015} or even modifying gene expression by stimulating ultrasound\cite{Hernot2008}. Moreover, cavitation-induced shock waves and associated shear forces, arising from imploding cavitation bubbles\cite{Akhatov2001}, could be employed for lithotripsy\cite{Yoshizawa2009} or controlled softening of calcified vessels\cite{Siegel1988} with spatially precise and localized mechanical stress gradients\cite{Bailey2003}. The SonoLancet could potentially serve as a way to induce streaming of biological fluids and translation of entities inside the body. Unidirectional motion of drugs or drug nanocarriers, induced by the transfer of momentum from the wave to the liquid, could enhance pharmaceutical effectiveness\cite{Mitragotri2005} and facilitate applications such as targeted and localized drug, stem cell or gene delivery. Atomization provides the ability to transform liquid into micro-droplets with control over droplet size distribution and direction of the droplet jets allowing new platforms for delivery of drugs or cells \textit{e.g}. within bodily cavities such as those in the respiratory system. 

Considering the safety, the diameter of the volume of tissue influenced by SonoLancet is comparable to that of a 11G needle (outer diameter = \SI{3.05}{mm}), while the energy is delivered through a small tract generated by a 21G needle (outer diameter = \SI{0.82}{mm}). This allows influencing a large volume of tissue in respect to the small size of the needle, while minimizing trauma, when accessing target site. Therefore, the SonoLancet could contribute to modifying the tissue permeability for drug delivery applications at precise tissue locations and with extreme ultrasound exposure, or allow minimally invasive surgical interventions \textit{via} spatially precise histotripsy. Importantly, in all of these experimentations the employed ultrasound frequencies were relatively low, < \SI{100}{\kHz}. Moreover, the risk of cavitation to occur and induce mechanical damage in unwanted regions inside tissue (\textit{e.g.} along the needle shaft) is minimized by the fact that the acoustic energy is highly concentrated at the tip of the needle, as supported by numerical simulations (Figure \ref{Figure1}B(1,2)). For this reason, cavitation is more likely to appear at the needle tip than elsewhere, and if such event were to occur along the needle shaft, its effects would be limited in a confined region from the needle boundary, as the intensity attenuates strongly in the direction radial to the needle center axis due to geometric factors. Since at these frequencies the attenuation coefficients in tissue and water are small, the heat deposition to tissue is minimized. This is supported by low temperature rise recordings < \SI{2}{\celsius} and in extreme conditions < \SI{4}{\celsius}. Therefore, these findings suggest that the SonoLancet provides a minimally invasive way to influence matter \textit{via} non-linear ultrasonic mechanisms within a confined volume at the proximity of the needle without major thermal impact. To our knowledge, this study is the first to demonstrate the selected NLU effects near the ultrasonically actuated medical needle. While this study exemplified the use of ultrasonically actuated needle in the biopsy application, its use could extend to other medical applications.

To conclude, we have functionalized the medical needle tip by nonlinear ultrasonics and, at a micrometer to millimeter scale, and quantitatively characterized its physical properties such as cavitation, acoustic radiation pressure and atomization within a confined volume defined as the SonoLancet. In this approach, the conventional hypodermic needle structure employs the lumen as a conduit to bring sound energy to the needle tip. The 3D shape of the conventional needle bevel amplifies the ultrasound action resulting in localized nonlinear effects at the proximity of the needle tip. Limiting the ultrasound action to the SonoLancet provides spatiotemporal control of sound-tissue interaction and with respect to minimizing microtrauma of adjacent tissues. By bringing precisely controlled ultrasound energy to a specific location in the body, the SonoLancet has the potential to give conventional medical needles a new role in many applications, such as tissue biopsy, molecular diagnostics, drug or gene delivery, cell modulation, or minimally invasive surgical procedures. 

\section*{Methods}

\subsection*{Experimental arrangement}

\begin{figure}[ht]
	\centering
	\includegraphics[width=\linewidth]{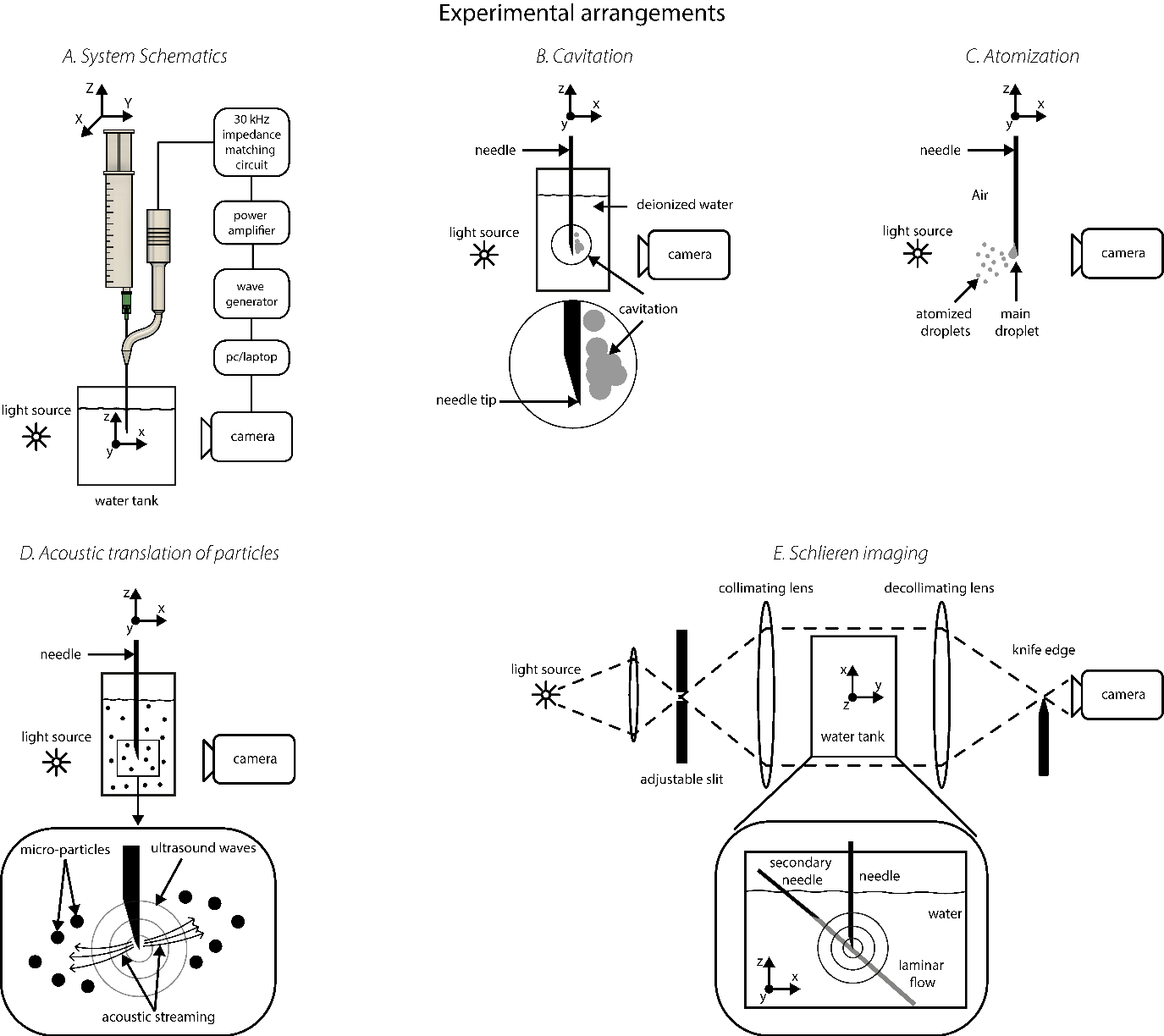}
	%\internallinenumbers
	\caption{Experimental arrangements: (A) Simplified schematics of the ultrasound system arrangement with optical detection of needle tip activity, (B) cavitation experiment, (C) atomization experiment, (D) acoustic streaming experiment and the (E) Schlieren imaging experiment.}
	\label{Figure6}
\end{figure}

A Langevin ultrasound transducer (Figure \ref{Figure1}A(1) and Figure \ref{Figure6}A) was coupled \textit{via} a custom-designed solid 3D printed aluminum waveguide (3D Step Oy, Ylöjärvi, Finland) to a 21G $\times$ \SI{80}{mm} hypodermic needle (model: 4665465, 100 STERICAN, B Braun, Melsungen, Germany) connected to a \SI{10}{mL} syringe (catalogue number: 12931031, Plastic PP Syringes Luer Slip, Fisherbrand, Fisher Scientific, Hampton, NH, United States) (Figure \ref{Figure1}A(1)). The device was powered by a computer-controlled combination of an RF amplifier (model: AG 1012LF, Amplifier/Generator, T\&C Power Conversion, Inc., Rochester, NY, United States) and a function generator (model: Analog Discovery 2, Digilent, Inc., Henley Court Pullman, WA, United States), which provided the driving signal. High speed (HS) imaging of needle actuation was employed to observe and quantify the NLU phenomena in different experiments. The HS camera (model: Phantom V1612, Vision Research, Wayne, NJ, United States) was operated in conjunction with a macro lens (model: Canon MP-E 65 mm f / 2.8 1-5x Macro Photo, Canon Inc., Ōta, Tokyo, Japan) and a general purpose lens (model: Canon EF-S 15-85 mm f / 3.5-5.6 IS USM, Canon Inc.). A white light-emitting diode (catalogue number: 4052899910881, White Led, 3000 K, 4150 lm, Osram Opto Semiconductors, Germany) was used as a light source to produce background light for shadowgraphy of the object during the HS camera recordings. The spatial position of the needle tip was controlled by a motorized three-axis translation stage (model: 8MT50-100BS1-XYZ, Motorized Translation Stage, Standa, Vilnius, Lithuania) operated by a computer user interface programmed in LabVIEW. Additionally, a combination of a dual-axis goniometer (model: GNL20/M - Large Dual-Axis Goniometer, Thorlabs, Inc., Newton, NJ, United States) and a continuous rotation stage (model: CR1/M - Continuous Rotation Stage, Thorlabs, Inc., Newton, NJ, United States) was mounted on the translation stage to manually control the polar and azimuthal angle of the needle shaft.

\subsection*{Numerical simulation}

In this study, the software COMSOL Multiphysics v5.5\cite{2020COMSOLV.5.5} was used to simulate the acoustic emission of the needle in a volume of water. The 3D geometry consists of an ultrasound transducer coupled to a 21G hypodermic needle \textit{via} an S-shaped aluminum waveguide (Figure \ref{Figure1}A(1)). The needle shaft is surrounded by a cylindrical volume of water (external dimensions = R $\times$ H = \SI{5}{mm} $\times$ \SI{50}{mm}). The ultrasonic actuation of the needle is obtained by applying a potential difference of \SI{100}{V} at a frequency of \SI{31.32}{\kHz} across the piezoelectric stack. The pressure field in water, generated by the vibrating needle, was calculated by performing a frequency domain analysis. The x-component of the time-average acoustic intensity, defined as $I_{ta,x}=\frac12 \operatorname{\mathbb{R}e}\{pv_x^*\}$ with p being the complex pressure and $v_x^*$ the complex conjugate of the acoustic velocity along the x-direction, was then evaluated on the xy-plane coincident with the needle center axis (Figure \ref{Figure1}B(1)) and on a line adjacent to the outer needle surface, starting from the needle tip and ending at the waveguide-needle attachment (Figure \ref{Figure1}B(2)).

\subsection*{Cavitation}

Cavitation experiments were performed in an acrylic chamber (external dimensions = L $\times$ W $\times$ H = \SI{21}{mm} $\times$ \SI{14}{mm} $\times$ \SI{15}{mm}, wall thickness = \SI{5}{mm}) filled with deionized water degassed to \SI{5.8}{\mg\per\liter} (catalogue number: 11754266, Traceable Portable Dissolved Oxygen Meter, Fisherbrand, Fisher Scientific, Hampton, NH, United States) (Figure \ref{Figure6}B). The needle was immersed to a depth of \SI{45}{mm} from the water surface and subsequently actuated by a continuous sinusoidal signal in the frequency range from \SI{31}{\kHz} to \SI{32}{\kHz} and emitted TAP of $\sim$ \SI{0.2}{W}. HS recordings were taken in order to visually capture the cavitation activity around the needle at different positions along the needle length. More specifically 45 videos (sample rate = \SI{300 000}{} frames per second (fps), exposure = \SI{2.8}{\us}, resolution = 128 pixels$\times$112 pixels, lens aperture = 3.3; lens model: Canon MP-E 65 mm) were acquired, covering a total length of \SI{45}{mm} from the needle tip to the waveguide after vertical concatenation. These settings permitted to achieve an image resolution of about \SI{9}{\um\per pixel}. The HS videos were finally analyzed by an algorithm implemented in MATLAB (R2020b)\cite{Mathworks2016} capable of tracking the needle position over time and quantify the cavitation activity. The needle displacement along the x-axis from its rest position was measured using sub-pixel image registration by cross-correlation in the frequency domain\cite{Guizar-Sicairos2008} between the $i^{th}$ frame $I_i(x,y)$ and the first frame $I_1 (x,y)$ of the high speed footage as follows:
%\begin{linenomath*}
\begin{equation}
\begin{aligned}
\Delta x_i = \argmax_x (FFT\{I_i (x,y)\} \cdot FFT\{I_1 (x,y)\}^*)
\end{aligned}
\label{disp}
\end{equation}
%\end{linenomath*}
where $\Delta x_i$ is the measured needle x-displacement from the rest position and $FFT$ is the two-dimensional fast Fourier transform. In order to quantify the projected area of cavitation a new set of frames, where only the cavitation activity is present, is created. This is done by subtracting the reference image $I_1$ (rigidally translated by $\Delta x_i$ in order to match the needle position in the $i^{th}$ frame) from each frame of the original footage:
%\begin{linenomath*}
\begin{equation}
\begin{aligned}
I_{cavitation,i}(x,y) = iFFT\{(FFT\{I_i(x,y)\}-(FFT\{I_1(x+\Delta x_i,y)\}\}
\end{aligned}
\label{cav}
\end{equation}
%\end{linenomath*}
where $iFFT$ is the two-dimensional discrete inverse Fourier transform. A probability map $P_{cavitation} (x,y)$, in which the value of each pixel represents the probability to observe a cavitation event in a specific coordinate, is then measured as follows:
%\begin{linenomath*}
\begin{equation}
\begin{aligned}
P_{cavitation}(x,y) = \frac{100}{N} \sum_{i = 1}^{N} I_{cavitation,bw,i} (x,y),
\end{aligned}
\label{prob}
\end{equation}
%\end{linenomath*}
where $N$ is the total number of frames and $I_{cavitation,bw}(x,y)$ is a frame-set generated by thresholding $I_{cavitation} (x,y)$ with the Otsu method\cite{Otsu1996}. The projected area of cavitation as a function of time is then obtained:
%\begin{linenomath*}
\begin{equation}
\begin{aligned}
A_{cavitation,i} = \int\int I_{cavitation,bw,i} (x,y) dx\kern 0.16667em dy
\end{aligned}
\label{cavarea}
\end{equation}
%\end{linenomath*}

Finally, a total of 45 probability maps were vertically concatenated to reproduce the entire length from the needle tip to the waveguide.

\subsection*{Atomization}
The purpose of this experiment was to demonstrate that with an ultrasonically actuated needle one can eject micro-droplets from the needle tip surface and control the droplet size distribution. A droplet of deionized water was introduced near the needle tip, while the needle was operated in air (Figure \ref{Figure6}C). HS recordings of the atomization process were acquired with the following settings: sample rate = \SI{16653}{fps}, exposure = \SI{10}{\us}, resolution = 768 pixels $\times$ 768 pixels, lens aperture = 16 (lens model: Canon MP-E 65 mm). The video frames were then analyzed in MATLAB to determine the droplet size distribution for different excitation frequencies and to compare it with theoretical predictions. Each frame $I_i (x,y)$ was first segmented with an adaptive thresholding\cite{Bradley2007}. Then, the built-in MATLAB function ‘regionprops’ was applied to measure the area and the equivalent diameter for each 8-connected component in the binary image $I_{bw,i} (x,y)$. Finally, the median droplet diameter distribution was calculated over $N = 1000$ randomly selected frames and compared with theoretical predictions defined by Equation \eqref{lang}.

\subsection*{Acoustic streaming}
Polystyrene based microparticles with a diameter of \SI{30}{\um} (catalogue number: 84135-5ML-F, Micro particles based on polystyrene analytical standard, size: 30 $\mu$m, Sigma-Aldrich, St. Louis, MO, United States) were used to visualize their translations near the needle tip. The test was conducted in a plastic cuvette (external dimensions = L $\times$ W $\times$ H = \SI{12}{mm} $\times$ \SI{12}{mm} $\times$ \SI{45}{mm}, wall thickness = \SI{1}{mm}) filled with deionized water (\SI{3}{mL}) with a particle concentration of 0.05\% v/v (Figure \ref{Figure6}D). A low power continuous sinusoidal signal (TAP = \SI{0.1}{W}) was applied to the needle, whose action was recorded with the following care settings: sample rate = \SI{10000}{fps}, exposure = \SI{2}{\us}, resolution = 1024 pixels $\times$ 800 pixels, lens aperture = 2.8 (lens model: Canon MP-E 65 mm). An averaged velocity field $\bar V (x,y) = \bar v_x \hat x + \bar v_y \hat y$, being $\hat x$ and $\hat y$ unit vectors, was then calculated by employing a cross-correlation technique\cite{Jambunathan1995}. Each frame $I_i (x,y)$ is divided into small interrogation boxes of the size of 64 pixels $\times$ 64 pixels. Considering $I_i (x,y)$ and $I_{i+1}(x,y)$to be the frames recorded at time $t$ and $t+dt$ , let $f(x,y)$ be an interrogation box centered at specific location $(x,y)$ in $I_i (x,y)$ and $g(x,y)$ an interrogation box centered at specific location $(x,y)$ in $I_{i+1} (x,y)$. The displacement field $\vec u (x,y)$ of two consecutive frames is calculated by measuring the offset $(\Delta x, \Delta y)$ of the peak of the correlation map between $f(x,y)$ and $g(x,y)$ as follows:
%\begin{linenomath*}
\begin{equation}
\begin{aligned}
&\Delta x_i = \argmax_x (FFT\{g (x,y)\} \cdot FFT\{f (x,y)\}^*),\\
&\Delta y_i = \argmax_y (FFT\{g (x,y)\} \cdot FFT\{f (x,y)\}^*),\\
&\vec u_i (x,y) = \Delta x_i(x,y)\hat x + \Delta y_i(x,y)\hat y.
\end{aligned}
\label{vel}
\end{equation}
%\end{linenomath*}

The averaged velocity map is then calculated as:
%\begin{linenomath*}
\begin{equation}
\begin{aligned}
\bar V(x,y) = \frac{1}{N} \sum_{i = 1}^{N} \frac{\partial}{\partial t} \vec u_i (x,y),
\end{aligned}
\label{avgvel}
\end{equation}
%\end{linenomath*}
where $N = 1000$ is the total number of frames. 

\subsection*{Schlieren imaging}

The experiment was performed in a glass chamber (external dimensions = L $\times$ W $\times$ H = \SI{250}{mm} $\times$ \SI{250}{mm} $\times$ \SI{250}{mm}, wall thickness = \SI{4}{mm}) filled with deionized and degassed water of temperature $T$ = \SI{22}{\celsius} as measured by a thermocouple thermometer (model: FLUKE T3000 FC, Fluke, Everett, WA, United States). A laminar flow of water (\SI{15}{\celsius}, $v$ = \SI{0.3}{\meter\per\second}) emanating from a secondary 21G needle directed towards SonoLancet (Figure \ref{Figure2}B and \ref{Figure6}E) was generated on a plane parallel to the needle center axis. Ultrasound continuous waves of \SI{31.7}{kHz} and a TAP $\sim$ \SI{0.8}{W} were applied to the needle for \SI{10}{s}. HS Schlieren recordings of the disturbance of a water laminar flow induced by the needle action were acquired (sample rate = \SI{100000}{fps}, exposure = \SI{0.357}{\us}, resolution = 128 pixels $\times$ 128 pixels, lens model: Canon EF-S 15-85 mm).

\subsection*{Tissue phantom experiment}

Experiments in a tissue phantom (10\% w/v porcine gelatin/deionized water, catalogue number: 53028, Gelatin from bovine and porcine bones for ballistic analysis Type 1, Honeywell Fluka, Morris Plains, NJ, United States) were carried out in combination with the ultrasonically actuated needle pre-filled with contrast agent (0.1\% w/v, fluorescent particles/deionized water, catalogue number: 11894932, Fluoro-Max Dyed Red Aqueous Fluorescent Particles, Thermo Fisher Scientific, Waltham, MA, United States). The needle was inserted to a depth of \SI{7}{mm} from the gelatin surface, penetrating three times the sample in order to let the contrast agent wet the walls of the created cavity. Continuous ultrasound wave at \SI{31.4}{kHz} and TAP $\sim$ \SI{0.8}{W} was applied to the needle for \SI{10}{s}. HS videos of projections of the volume of interaction between the gelatin and the SonoLancet were taken using the following settings: sample rate = \SI{1000}{fps}, exposure = \SI{10}{\us}, resolution = 768 pixels $\times$ 768 pixels, lens aperture = 16 (lens model: Canon MP-E 65 mm). Optical projection tomography scanning of the sample was performed using an OPT scanner (Bioptonics OPT 3001M Scanner). Reconstructed images (resolution = 512 pixels $\times$ 512 pixels), which represent the intersection between planes perpendicular to the needle center-axis and the volume of actuation in the tissue phantom, were generated using NRecon version 1.6.1.0 (Skyscan) software and further analyzed in MATLAB. Each reconstructed image $I_z (x,y)$ was first thresholded with the Otsu method. A sequence of morphological erosion and dilation using a disk-shaped structuring element of radius $r=\SI{10}{pixels}$ was then applied to fill eventual gaps in the image and remove small unwanted objects arising from noise. Contours of the sections, defined as sequences of n contour points $\{P_0,P_1,…,P_{n-1}\}$, are extrapolated from the images and filtered with a Savitzky-Golay finite impulse response (FIR) smoothing filter (polynomial order = 2, window size = 11). Finally, the 3D surface of actuation around the needle tip is constructed with triangular elements and further adjusted in Fusion 360. 

\subsection*{USeFNAB}

We quantified the ultrasound-tissue interaction by measuring the tissue yield of ultrasound-enhanced fine-needle aspiration biopsies obtained from bovine liver, kidney, spleen and muscle at 4 different TAP levels (\SI{0}{W} control; sonication groups: \SI{0.2}{W}, \SI{0.5}{W}, \SI{0.8}{W}) with the yield obtained with the conventional fine-needle aspiration biopsy approach (FNAB control) having the same syringe (\SI{10}{mL}) and needle (21G $\times$ \SI{80}{mm}) as USeFNAB. Liver (from 33 months old female bovine), kidney (from 33 months old male bovine), spleen (from 55 months old female bovine) and striated muscle (diaphragm, from 19 months old female bovine) specimens were obtained from the slaughterhouse (Vainion Teurastamo Oy, Orimattila, Finland) within \SI{1}{\hour} \textit{post mortem} and experiments were performed 3-\SI{5}{\hour} \textit{post mortem} at room temperature (22-\SI{24}{\celsius}). Only peripheral parts of the same liver were used, to avoid large blood vessels and acquisition of bloody samples. In liver and in spleen, samples were taken at a depth of \SI{2}{cm} from the organ surface by keeping the needle parallel to the table, where the target tissue is fixed throughout the procedure. Sample acquisition in kidney was limited to the cortical area and each biopsy was performed in different renal lobes, while, in muscle, biopsies were taken along the fibre orientation. The used needles and syringes were weighed before and after the biopsies using a balance with resolution of \SI{0.1}{mg} (model: Quintix Precision Balance 5,100 g $\times$10 mg, Sartorius AG, Germany) to measure the mass of the sample extracted from the tissue. The test was divided into 6 sets of 5 repetition. In each set, the order of the groups to be tested (FNAB (\textit{n} = 6); USeFNAB: \SI{0}{W} (\textit{n} = 6), \SI{0.2}{W} (\textit{n} = 6), \SI{0.5}{W} (\textit{n} = 6) and \SI{0.8}{W} (\textit{n} = 6)) was randomized with a random permutation within a set of five samples (one sample from each group). The FNAB procedure consisted of inserting the tip of the needle into the target tissue using a ‘fanning technique’, pulling the plunger back to the \SI{2}{mL} mark and moving the needle back and forth 7 times with a frequency of 1 pass per second. The penetration depth was approximately \SI{15}{mm} at a targeted angular offset of approximately \SI{5}{\degree} between the passes. The suction was maintained manually throughout the biopsy sampling by keeping the plunger of the syringe at the the \SI{2}{mL} mark and only released after the needle was withdrawn\cite{Kreula1989}. The same procedure was adopted for USeFNABs, where ultrasound continuous waves (f = \SI{31.7}{kHz}, pulse repetition frequency (PRF) = \SI{55}{Hz}, duty cycle (DC) = \SI{52.05}{\%}) were applied throughout the biopsy after the needle was inserted into the tissue. The selected parameters of the protocol were considered appropriate to obtain a sufficient amount of sample for yield analysis or to be processed into agarose cell blocks. The operator skills in performing the biopsy and the protocol were confirmed by a specialist in radiology (J.K.) with considerable experience in FNAB. Assuming that normal distribution does not apply, the FNAB mass was compared to the masses of USeFNAB groups using Mann–Whitney U test. The Bonferroni-corrected level of statistical significance was 0.0125.

\subsection*{Histology}
Biopsy samples were processed into agarose cell blocks for histological assessment\cite{Koh2013}. Aspirates were expressed into \SI{1.5}{mL} tubes (11569914, Snap Cap Low Retention Microcentrifuge Tubes, Thermo Fisher Scientific, Waltham, MA, United States) containing \SI{500}{\uL} of 1x PBS (BP399-4, Phosphate Buffered Saline, 10X Solution, Fisher BioReagents, Fisher Scientific, Hampton, NH, United States) and subjected to centrifugation at \SI{500}{G} for \SI{5}{\minute}. After a second PBS washing, pellets were resuspended in \SI{500}{\micro\liter} of 10\% neutral buffered formalin (15817114, HistoTainer, Simport, Saint-Mathieu-de-Beloeil, Quebec, Canada). After \SI{10}{\minute} of incubation time, tubes were centrifuged at \SI{500}{G} for \SI{5}{\minute}, then, the supernatant was carefully discarded, and pellets were washed in PBS twice. Excess PBS was carefully removed and formalin fixed pellets were resuspended in \SI{100}{\uL} of molten agarose (10377033, Agarose Low-Melting, Nucleic Acid Recovery/Molecular Biology Grade, Thermo Fisher Scientific, Waltham, MA, United States) kept at \SI{42}{\celsius} and gently centrifuged at \SI{180}{G} for \SI{1}{\minute}. The agarose cell blocks were let solidify at \SI{4}{\celsius} and then transferred into 70\% v/v ethanol (172381, 70\% v/v Ethanol Solution Molecular Biology Grade, Fisher Scientific, Hampton, NH, United States) pre-filled tubes. Several sections of the paraffin embedded cell blocks (4 to 5 for each sample) were cut at \SI{5}{\um} thickness, mounted on \SI{25}{mm} $\times$ \SI{75}{mm} slides (J1800AMNZ, SuperFrost Plus Adhesion slides, Thermo Fisher Scientific, Waltham, MA, United States) and stained with hematoxylin and eosin (H\&E). Finally, microscopy pictures (bright field, 20x magnification, extended focus) were generated using 3DHISTECH Panoramic 250 FLASH II digital slide scanner. Four biopsy samples (liver, \textit{n} = 1; spleen, \textit{n} = 1; muscle, \textit{n} = 2) failed to be processed for histological analysis due to breakage of the agarose cell block.

\subsection*{Quantification of sample fragmentation}
\begin{figure}[ht]
	\centering
	\includegraphics[width=\linewidth]{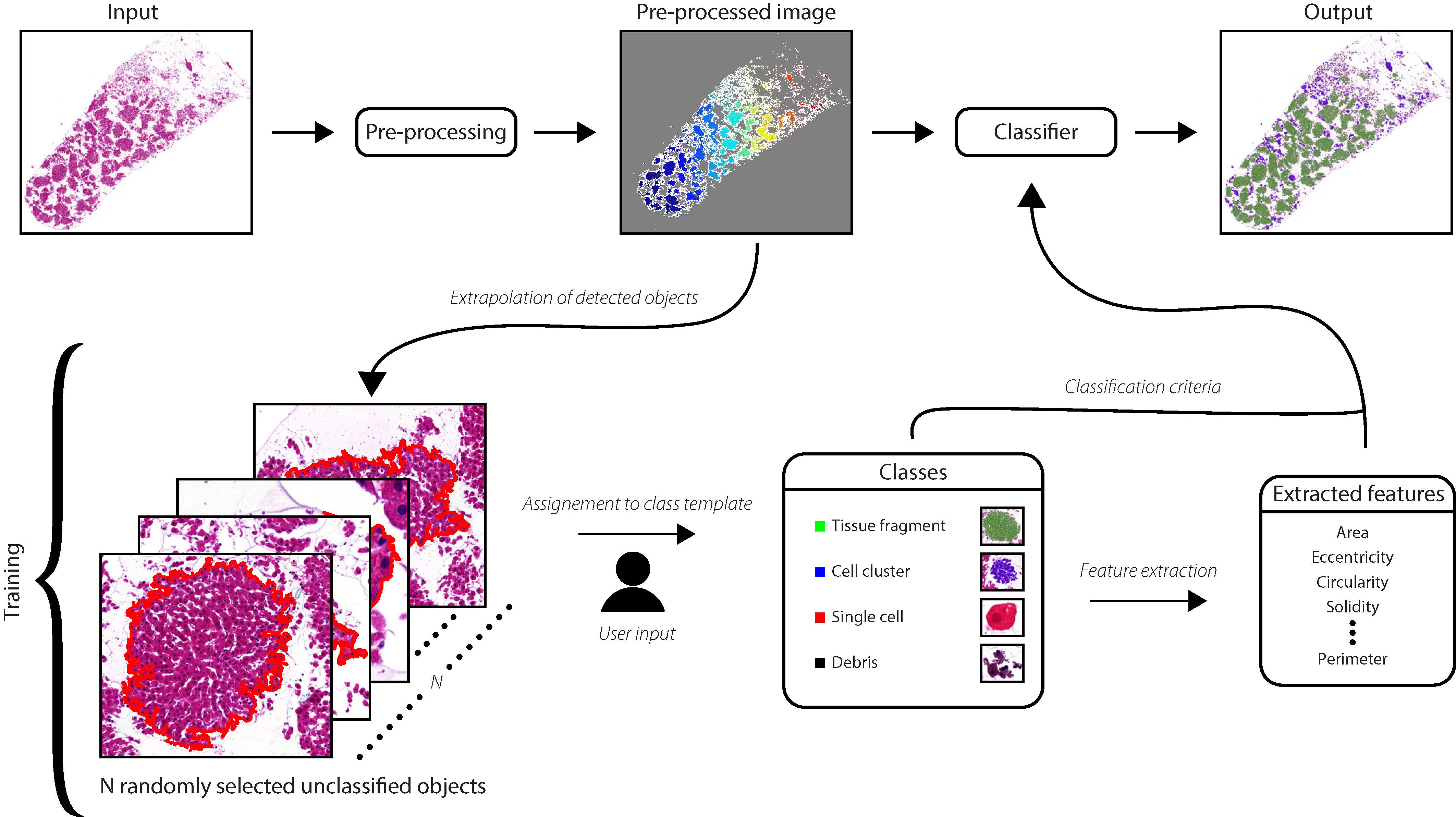}
	%\internallinenumbers
	\caption{General description of the classification process. During the training phase a number N of randomly selected unclassified objects objects are selected from the data and firstly assigned to one of the output classes (tissue fragment, cell cluster, single cell, debris) by a user. Several morphological features (area, eccentricity, circularity, \textit{etc.}) are then extracted from such objects and used as classification criteria during the classification phase.}
	\label{Figure7}
\end{figure}

A supervised classification method has been implemented in MATLAB to quantity the fragmentation of the liver samples by means of feature extraction of objects identified in the histological images\cite{Kumar2015}(Figure \ref{Figure7}). The classification algorithm is able to detect four different histological structures, namely tissue fragments (large regions of adjacent hepatocytes), cell clusters (small aggregation of hepatocytes), single cells and debris (organic waste). The dataset is composed of 29 histological slides belonging to the following sample groups: FNAB (\textit{n} = 5); USeFNAB: \SI{0}{W} (\textit{n} = 6), \SI{0.2}{W} (\textit{n} = 6), \SI{0.5}{W} (\textit{n} = 6) and \SI{0.8}{W} (\textit{n} = 6). Each image $I_h(x,y)$ was first converted into gray-scale image and then subjected to thresholding using the Otsu method. The the Moore-Neighbor tracing algorithm modified by Jacob's stopping criteria\cite{Gonzalez2004} was applied to trace the boundaries of all 8-connected objects detected in the binary image $I_{bw,h}(x,y)$. The objects were assigned to a certain class based on a similarity score method defined as follows:
%\begin{linenomath*}
\begin{equation}
\begin{aligned}
&C = \argmin_k S(x_i,k),\\
&S(x_i,k) = \sum_{m = 1}^{M}\frac{|x_{i,m}-b_{k,m}|}{c_{k,m}},\\
&b_{k,m} = \frac1N\sum_{n = 1}^{N}Y_{k,m,n},\\
&c_{k,m} = \sqrt{\frac1N\sum_{n = 1}^{N}(Y_{k,m,n}-b_{k,m})^2}.\\
\end{aligned}
\label{simscor}
\end{equation}
%\end{linenomath*}
In the above, $C$ is the the output class which each object is assigned to by the classifier, $x_i$ is a vector containing a number $M$ of morphological features (\textit{i.e.} area, eccentricity, circularity, solidity, \textit{etc.}) extracted from the $i_{th}$ object and $k$ is an integer number corresponding to a certain class (1 = tissue fragments; 2 = cell clusters; 3 = single cells; 4 = debris). $S(x_i,k)$ is the score function whose output value denotes how well an object is represented by a class $k$ (the lower the score, the higher the similarity). $Y_{k}$ is a $M\times N$ matrix containing feature vectors belonging to pre-classified objects that were priorly assigned to a class $k$ in the supervised learning phase. Finally, several indicators of sample fragmentation (\textit{i.e.} total number of tissue fragments, cell clusters, single cells, debris pieces and their respective average area) were extracted and used for statistical comparison between the FNAB and USeFNAB groups. Assuming that normal distribution does not apply, the Mann–Whitney U test was chosen, with a Bonferroni-corrected level of statistical significance of 0.0125.

\bibliography{library}

\section*{Acknowledgements}

We thank Mr. Matti Mikkola, M.Sc., Mr. Yohann Le Bourlout, M.Sc., Mr. Pauli Tuovinen, B.Sc. Nallannan Balasubramanian, Ph.D. and Adjunct Prof. Gösta Ehnholm, Ph.D., for constructive discussions related to the topic, and Dr. Nick Hayward, MBBS, Ph.D. for linguistic review of the manuscript. We thank Jyrki Kreula, MD, Ph.D. for guidance and providing us extensive information related to the FNAB procedure. We also would like to thank Prof. Kari Eklund, MD, Ph.D. for support in histological experimentation. We thank the Tissue Preparation and Histochemistry Unit in the Department of Anatomy, University of Helsinki. Histology images were generated using the 3DHISTECH Pannoramic 250 FLASH II digital slide scanner at Genome Biology Unit supported by HiLIFE and the Faculty of Medicine, University of Helsinki, and BioCenter Finland. The Academy of Finland is acknowledged for financial support (grants 314286 and 311586). 

\section*{Author contributions statement}

All authors contributed to the design of the study, writing or reviewing the manuscript and have approved the final version of the manuscript. Emanuele Perra produced all data and conducted all data analysis, except Kenneth Pritzker and Emanuele Perra also analyzed the histology images. Emanuele Perra was responsible for the histology process.

\section*{Conflict of Interest}

 Heikki Nieminen, Kenneth Pritzker, Edward Hæggström and Eetu Lampsijärvi have stock ownership in Swan Cytologics Inc., Toronto, ON, Canada and are inventors within the patent application WO2018000102A1. Emanuele Perra, Gonçalo Barreto, Muhammad Arif and Tuomas Puranen do not have any competing interests in relation to this work. 

\section*{Data availability}

The datasets are available upon request.

\section*{Code availability}

The codes are available upon request.

\end{document}